# IS OUR RESEARCH PRODUCTIVITY IN DECLINE?
## A New Approach in Resolving the Controversy

Gennady Shkliarevsky

**Abstract:** This contribution examines the current controversy over research productivity. There are two sides in this controversy. Using extensive data from several industries and areas of research, one side argues that research productivity is currently in decline. The other side disputes this conclusion. It contends that the data used in making this argument are selective and limited; they do not reflect the overall state of research. The conclusion that follows from this critique is that the indicators of research productivity we currently use are not reliable and do not warrant a definitive answer to the problem.

The article agrees that we need a new set of indicators in assessing research productivity. It proposes that we should look at global indicators related to knowledge production in general, rather than look at selective data that are inevitably limited in their scope. The article argues that the process of creation plays the essential role in knowledge production. Therefore, the perspective that uses the process of creation as its central organizing principle offers a unique and global view on the production of knowledge and makes a definitive resolution of the controversy possible. The article also outlines some steps for improving research productivity and realizing the full potential of the human capacity to produce knowledge.

## Introduction

Since the onset of the industrial age the world has experienced an unprecedented progress that has transformed our civilization. Much of this progress is due to systematic and concerted efforts to expand our knowledge. The logic of these efforts is very straightforward. It follows from the widespread belief that investment is the key to stimulating knowledge production. Government agencies and private investors across the globe have contributed billions upon billions to stimulate knowledge growth, expecting that these contributions will lead to new ideas and approaches that will lead to the continued and sustained advancement of our civilization.

Investments in knowledge production have been increasing over the years. Allocating more resources to research and development is a popular theme these days. Many regard such allocations as a major, if not the only way to solve the problems that the world currently faces. This expectation, for example, has recently moved American Congress to increase dramatically the funding for research and development in the U.S. as a way of moving the country forward.[1]

Yet the wisdom of such investments does not go unchallenged. There is a growing sense that the returns on these investments have been progressively and even exponentially diminishing. Many researchers point out to abundant data showing that research productivity over the last several decades has been in decline. They differ in their explanations for this development but they are uniform in their pessimistic conclusion.

There are critics who challenge this conclusion. They dispute the data and question approaches used in analyzing it. However, they do not definitively reject the conclusion. Their main point is that currently there is simply no effective way of measuring innovation. In other words, they argue that we simply do not know whether our research productivity is in decline or not. The critics do not claim that the controversy over research productivity is in principle irresolvable, which suggests that a different approach may be successful in providing its definitive resolution.

This article takes up the challenge of resolving the controversy over research productivity. It explores a different route in addressing this problem. Rather than focus on micro facts from different areas of research and development and argue for or against their reliability, it proposes to look at the facts that are more general: the patterns of the process that generates knowledge in general and the effects that new knowledge produces on our global view of reality as a whole, as well as on our use of resources, including environment. The pursuit of this route will require a discussion of what knowledge is and how it is produced. The discussion will provide the answer to the critical questions why and how knowledge grows and why it grows exponentially. Just like the current approaches in this controversy, the approach proposed in this article is also fact based. However, unlike the facts used by the current contributors to the debate, the facts the new approach uses are not a subject of dispute and, therefore, offer a prospect for definitive resolution.

**The Controversy Over Research Productivity**

For over several decades there has been a steady stream of contributions in both scholarly and popular venues that alert the public to a decline in research productivity. Perhaps the biggest concern expressed in these contributions is that they do not see this decline as a temporary development but as a long-term tendency that has persisted for decades with significant consequences for our economy and society.[2] The titles of these contributions are alarming. Several examples give a good idea of their general tone: "Diminishing Returns? U.S. Science Productivity Continues to Drop,"[3] "Slowed Canonical Progress in Large Fields of Science,"[4] "Bang for the R&D Buck Is in a Long, Steady Decline,"[5] "New Ideas Are Getting Harder to Find--and More Expensive,"[6] "In an Era of Tech Innovation, Whispers of Declining Research Productivity."[7] These are just some of the milder ones. The title of the article "We Are Out of Big Ideas"[8] that appeared in *The Wall Street Journal* is much more dramatic and looks like a cry of desperation.

Summarizing what these and many similar contributions[9] convey to the public, Philip Böing and Paul Hünermund write in their article "A Global Decline in Research Productivity? Evidence from China and Germany":

> Economists have proposed that continual decline in research productivity at the technology frontier potentially drives the observed stagnating or slowing growth rates in advanced economies over time (Gordon, 2016; Cowen, 2011).[10]

The publication that has become a definitive contribution on this subject is the article that was published in 2017 by Nicholas Bloom and his co-authors and has attracted much attention. The article challenges one key assumption made by many so-called endogenous growth models that "a constant number researchers can generate constant exponential growth."[11] Using a metric of "research productivity" they have devised, Bloom and his co-authors examine a number of areas where growth heavily depends on research and knowledge. One example they provide relates to computing technology and the so-called Moore's Law.[12] Their findings show that the number of researchers needed to sustain Moore's Law has increased 18-fold since 1971 and, consequently, productivity per researcher has fallen in the same proportion. They have also looked in other directions to confirm their findings: agricultural production, medical research, mortality and life expectancy, new molecular entities, and the aggregate figures for the economy as a whole. Their results show that

> Taking the U.S. aggregate number as representative, research productivity falls in half every 13 years—ideas are getting harder and harder to find. Put differently, just to sustain constant growth in GDP per person, the U.S. must double the amount of research effort searching for new ideas every 13 years to offset the increased difficulty of finding new ideas.[13]

The conclusion that intellectual productivity is exponentially declining everywhere has gained wide acceptance.[14] In less than five years since its publication, the article received 56- citations.[15] The prevailing view today is that research productivity is in decline, and not just in some individual countries, but rather that this decline has a global character.[16]

Despite this broad agreement, there are significant differences among those who make this claim. They disagree, for example, in their explanations of the decline. Some see it as the result of persistent and structural factors; others attribute it to transitory and fortuitous conditions. Although these disagreements present opportunities to have a multi-dimensional view of the decline, they do create confusion in assessing the magnitude of the problem.

Benjamin Jones sees structural causes of the decline. In his influential and much cited paper "The Burden of Knowledge and the 'Death of the Renaissance Man': Is Innovation Getting Harder?" argues that the cause is the very success in knowledge production. Due to the enormous accumulation of knowledge, generating new ideas requires a much longer period of education. One must clime on the shoulders of giants, which takes up considerably more time than in the past.[17] "If a rising burden of knowledge is an inevitable by-product of technological progress," Jones concludes, "then the model indicates pessimistic predictions for long-run growth."[18] Although Jones's prognosis does not offer much room for optimism, he does not exclude a possibility that the balance of forces affecting research productivity may change in the future, but the exploration of new possibilities, Jones insists, requires more investment needed. Chris Callaghan also points to a growing body of evidence that supports predictions of burden of knowledge models. According to Callanan, "We find some evidence supporting burden of knowledge perspectives." Like Jones, he all sees the growing need for more researchers and increased funding for research in the future.[19]

Among popular structural reasons for decline is one that emphasizes the lack of genuine competition of ideas and the general ossification of science. Salman Yousuf Guray and his co-authors observe, for example, a significant decline in productivity of medical faculty after attaining the rank of a professor.[20] Flora Tien and Robert T. Blackburn also see the relationship between the rank system and researchers' productivity.[21] Politization of research[22] and the dominance of hierarchies in the research community[23] are two other structural factors that a number of researchers consider an important cause of decline, although the impact of these factors still remains underinvestigated.

Then there are also many explanations that look away from structural causes to more passing and fortuitous conditions. Some researchers in this vein argue that the decline in knowledge growth is due to the graying of our academy. They explore the connection between the life cycle and research productivity.[24] Timothy Salthouse, for example, examines consequences of age-related cognitive decline on researchers' careers.[25] Other researchers find such claims objectionable;[26] and still others object to their objections.[27]

Laura Hunter and Erin Leahey insist that the demands of parenting have a serious negative effect on research productivity, particularly among female researchers.[28] Workaholism and the tendency to work long hours among researchers is also among popular explanations,[29] as is the impact of the pandemic.[30]

As one can see, there is no shortage of explanations for the decline in research productivity. They are very diverse and the only thing that they have in common is that they all accept the decline in knowledge growth as an indisputable fact. Although this diversity reveals many different aspects of the problem, they do take away from the general claim: with so many different and often contradictory explanations for the same phenomenon, one begins to wonder about the authenticity of the general claim they are supposed to affirm. Discords in explanations for the decline in research productivity have encouraged the dissent from the orthodox claim.

Indeed, dissonant voices have not been long in waiting. Although Erik Brynjolfsson and Andrew McAfee[31] largely agree that research productivity is in decline, their take on this development is different from that of the "burden of knowledge" theorists. In contrast to Jones and others, Brynjolfsson and McAffee argue that the declining rate of growth is relative, not absolute. They use data from digital industry to make their point. During the initial stage, innovative ideas did not require a great deal of effort: opportunities for innovation were abundant and researchers did not need to spend a great deal of time and effort to pick up what was easily accessible. However, as the industry advanced, in order to generate new ideas became, researchers had to go after innovations that required a great deal of more time and effort. Brynjolfsson and McAfee use a metaphor of a low growing and high growing fruit to explain their point. Researchers first went after low hanging fruit that was easy to pick. However, as research progressed they had to go after less accessible rewards.

In addition, Brynjolfsson and McAfee also argue that the character of research has changed over time. Advances in knowledge today, they claim, are largely due to small incremental changes that result from recombination of existing ideas, rather than generating radically new knowledge. The true work of innovation today, they write, "is not coming up with something big and new, but instead recombining things that already exist."[32] Digital innovation is, in their view, a perfect example of this progression to recombinant innovation.

For Brynjolfsson and McAfee, these observations serve as the basis for optimism. They indicate that knowledge growth is by far not over. Brynjolfsson and McAfee underscore that the potential for knowledge growth remains very high but we need to adopt new practices that befit the current stage. Knowledge growth, they claim, "is just being held back by our inability to process all the new ideas fast enough."[33] In other words, it is in our power to increase research productivity, or as another researcher put it, "Yes, Ideas Are Harder to Find. Don't Panic Yet."[34] All we need to do is to develop the capacity to process new ideas faster. Responding to their own rhetorical question whether knowledge growth is over, Brynjolfsson and McAfee emphatically proclaim: "Not a chance!"

Perhaps the most serious questioning of the entire issue of the existence of decline in research productivity has come from Alexey Guzey in the form of the response to Bloom under the title "Issues with Bloom et al's 'Are Ideas Getting Harder to Find?' and why total factor productivity should never be used as a measure of innovation."[35] Guzey sees several issues with the claim by Bloom and his co-authors. His main concern is that there is simply no reliable way to measure the global trend in research productivity. The total factor productivity (TFP) is not, in his view, a reliable way of measuring innovation. He also argues that selection of data from individual industries and research venues cannot substantiate conclusions regarding global trends in research productivity. His answer to the question whether research productivity declines is simple "We don't really know."

Following his criticism, Guzey maintains that there is a need to develop a more comprehensive and global approach to the issue of research productivity. He also insists that reforming research institutions and practices, rather than pouring more money into research, is a better path toward enhancing knowledge growth.[36]

Several takeaways that emerge from discussions of research productivity are relevant for the purposes of this article:

1. The main argument that the current research productivity is in decline suggests that knowledge production, just like any other form of production, also requires resources and these resources can be depleted.

2. The argument made, among others, by Brynjolfsson and McAfee that innovation today results primarily from recombination of the existing ideas points to one important conclusion.  Recombination of different ideas involves equilibration.  Therefore, equilibration is an important aspect in knowledge production.

3. Finally, the point made by Guzey, among others, that there is at this time no adequate metrics for assessing research productivity suggests that we need a new and more global approach toward knowledge production and that the current research institutions and practice are in need of reforms.

**Knowledge and the Process of Creation**

Discussions on research productivity have been going on for quite some time but even the most obvious questions related to this issue remain unanswered.  Researchers show that they may have some more or less informed intuitions about the dynamics of the current research productivity but they have not proven them conclusively and convincingly one way or another.  In this sense, one has to agree with Guzey that we really do not know whether research productivity declines or not.  Without a definitive answer, all theories on this subject are little more than intuitive approximations.  Indeed, there are many contributions on how to improve productivity, but without a fundamental theoretical insight into the production of knowledge, they offer little more than ad hoc recommendations.[37]  These contributions are mostly about pragmatic recommendations rather than in-depth explanations of the process involved.  At the end of the day, one must recognize that the only conclusion that has emerged from these discussions is that their results are inconclusive.  The subject of knowledge production is too deep and too important to be handled casually and without a theoretical foundation.

The failure to explain or even determine whether research productivity is in decline makes Guzey's argument regarding the need for a broader view increasingly attractive.  Indeed, approaching the problem of knowledge production from a different and more comprehensive global angle may be a fruitful alternative to the current focus on micro data.  One possibility for moving forward that this article sees is to look at the general process of knowledge production that is common to all spheres where knowledge is produced and to examine factors and conditions characteristic for the process as a whole.  Also, looking at major effects of knowledge production in cognate spheres—such as economy or use of resources, both human and non-human—may also provide useful indicators for determining knowledge growth.

*What is knowledge?*

Since this article deals with the production of knowledge, defining knowledge is a good first step on this challenging path of exploration. There is an abundance of definitions of knowledge that fit into different philosophical traditions. Yet despite this abundance, the definition of knowledge still remains elusive.[38] Even a very concise list of sources that discuss knowledge and its definitions will make a sizable list of entries.[39] A detailed examination of these definitions is certainly beyond the scope of this article. But a brief discussion is certainly in order for the purposes of this contribution.

Perhaps the most widely accepted definition is that knowledge is a justified true belief. Paul Moser, for example, offers the following concise characterization:

> Epistemology in the Western philosophical tradition has until recently offered a prominent definition of knowledge that analyzes knowledge into three essential components: justification, truth, and belief.[40]

He further adds that for a true belief to be knowledge, it must also have empirical confirmation—that is, an entity or a phenomenon in the real world that corresponds to this belief.[41] These basic features of the definition of knowledge appear in multiple other sources.[42]

A brief examination of these characteristics warrants the following conclusions. The assertion that knowledge is a form of belief means that it originates in the irrational sphere and that irrationality is a necessary aspect of knowledge. The specification that this belief must be true indicates that one must be very sincere in this belief and not treat it casually. The reference to justification reveals another important feature of knowledge. Justification refers to a logical operation that establishes the necessity of such belief. As an operation, justification is a form of equilibration. Consequently, this requirement indicates a general agreement that knowledge and, consequently, knowledge production involve equilibration. The reference to empirical verification spells out another important requirement of knowledge: there should be one-to-one correspondence established between justified true belief and external environment.

Although these characteristics of knowledge are widely accepted, they are not unproblematic. One can easily see that the view of knowledge as a form of belief is in tension with the insistence on rational justification of this belief—a tension that begs for an explanation. How can these two characteristics that come from different domains—one rational and the other irrational—be compatible with each other? The requirement of a one-to-one correspondence as a criterion for something to be regarded as knowledge is also problematic. As the history of science shows, many theories that passed the test of empirical verification have eventually turned out to be false or incomplete (the geocentric theory is a good example that withstood numerous empirical tests, yet proved to be limited). This brief analysis suggests that an explanation of the production of knowledge must demonstrate how these contradictory features complement each other within the same process.

*Evolution and Knowledge*

According to the current definition of knowledge, knowledge originates in the sphere of the irrational. Therefore, the process that gives rise to knowledge transcends reason and rationality—the two characteristics that we habitually associate with human knowledge. The conclusion that follows from this observation is that the process that creates human knowledge existed prior to the rise of the human mind and knowledge. The human mind did not create this process but rather this process created the human mind. Consequently, human knowledge is a product of the evolution that had taken place prior to the rise of humanity and in some form knowledge is characteristic for non-human reality. This realization is what has led researchers in their efforts to locate knowledge and cognition in non-human nature. Eshel Ben Jacob and his co-authors, for example, explore the roots of cognition in the world of bacteria.[43] However, considering the fact that the evolution is a cosmic phenomenon, the process that is involved in the creation of knowledge must reach even to non-life and should be conceived in very broad terms that transcend to boundaries of human existence.

Human knowledge involves symbolic operations. The level of mental organization that sustains symbolic operations is a product of the evolution. Obviously, the process that makes this evolution possible is the same process that is involved in the production of knowledge. Since the evolution operates on the scale of the entire universe, the process that propels the evolution also operates on the universal scale.

The evolution is primarily about the creation of new levels of organization. A level of organization is an integrated network of structurally connected operations that has its own distinct level of combinatorial power—that is, it offers a certain distinct set of possibilities, or degrees of freedom. We can identify many levels of organization that sustain certain kind of operations. For example, there are some levels of organization that sustain interactions among inanimate entities, such as particles or atoms; others sustain life and biological organisms; and there are levels of organization that sustain neural circuits and operations; and there are levels that sustain symbolic operations that constitute the human mind.

Each level of organization marks an advance in the evolution. Each of them represents a radical novelty, that is, it represents something that did not exist prior to its emergence. Each level of organization emerges from another that has preceded it and gives rise to another level that follows. For this reason, each level is more powerful than the one from which it has emerged. Since the emerging level of organization cannot be reduced to the level from each it has emerged (due to its superior power), we identify the act of its emergence as creation—an action or process of bringing something into existence.[44]

*The Main Function of the Process of Creation*

Power constitutes the main difference between the emerging level of organization and the level of organization from which it emerges. For example, the level of organization that sustains mental operations is more powerful than the level of organization that sustains different sensory-motor operations—visual, audio, tactile, olfactory, and gustatory. Mental images offer the same properties as do sensory-motor functions, but they also offer additional possibilities. For example, they allow having experiences that combine the experiences of sensory-motor functions. They even offer possibilities to experience sensations even when there is no external stimulus to activate them.[45] Moreover, mental images open the path to symbolic operations that offer unlimited possibilities for mental constructs.

How do new properties emerge? Due to power differential, there is simply no way that the level of organization that gives rise to the new level of organization can "know" or "design" the emerging level. Also, there can be no super-ordinate "designer" because there is no evidence of the existence of such "designer." And yet, the problem of design exists. The only way to resolve this problem is view it from a different angle. One must look at the conditions that create this problem. The only way we can expect there to be a "designer" is by thinking that the emergence of the new level of organization is the main function of the process that leads to this emergence. Just because one level of organization emerges from another, there is no reason to suppose that the rise of the new level of organization is the main function of the process that leads to its emergence. If we see the new level as merely a by-product of the more important function, the problem of "designer" simply disappears.

Since new and more powerful levels of organization sustain properties that exist at the levels of organization from which the former have emerged, we have to conclude that conservation is one important function of this process. Moreover, since operations are conserved as parts of combinations, they are activated more often, which also additionally helps to conserve them. Therefore, the conclusion that follows is that conservation is the main function in this process, and the emergence of new levels of organization is its result—a by-product of conservation.

The connection between the process that creates new and increasingly more powerful level of organization and conservation does not come as a surprise. After all, conservation is ubiquitous in nature. It is the essential aspect of our universe—one that makes its existence possible. Our universe is all there is. There is nothing outside the universe. Nothing can come into our universe and nothing can disappear from it because there is nowhere to disappear. Everything must be conserved. Conservation requires resources. Combinations that emerge in the course of conservation offer new possibilities and, thus, access to new resources. This process sustains the universe. It offsets the inevitable depletion of resources that is dues to entropy production. This process cannot bring the level of entropy production below zero—this is prohibited by the Second Law of Thermodynamics—but it can maintain this level at zero, which is not prohibited.

Since conservation creates new levels of organization, it leads to the evolution. What is evolution if not a succession of increasingly more powerful levels of organization nested in each other *matryoshka s*tyle. This process and the evolution it generates sustain our universe. The human mind is also a product of this process and, as such, has inherited its main properties and patterns.

Operations are a form of action. Being enacted conserves action. The more often an operation is enacted, the better it is conserved. Therefore, the creation of combinations is essential for conservation. The key to conserving an operation or an entity is to establish as many connections as possible. In order to be conserved, each operation sustained by a particular level of organization forms as many connections and combinations as possible. For $n$-number of operations at a given level of organization the number of possible combinations is $n^2$. In other words, if combinations represent an increased number of possibilities, or degrees of freedom, then the power of the new level of organization will represent an exponential increase.

The process of combining properties of individual operations is a form of equilibration. The equilibration of different operations creates a new and more powerful level of organization, that is, it creates disequilibrium. Conservation of this new level requires its equilibration with the operations sustained by the level of organization from which it has emerged, which means that these operations should adapt to the new level. The adaptation results in differentiation of the new level. Conservation of the new arrangement requires further equilibration, which opens a new cycle in the evolution. Thus, equilibration and the production of disequilibrium work together in a complementary mode and thus advance the overall evolution.[46]

The perspective that views knowledge and knowledge production through the prism of the process of creation explains how different features of the current definition of knowledge complement each other. It shows that the process that generates knowledge has its source in the sphere of the irrational. It also shows that the process of creation of knowledge requires equilibration. As forms of equilibration, logical operations and rational justification are integral to this process as emphasized by the current definition.

According to this perspective, the human cognitive structure includes many levels of organization—from biological to neural and sensory motor, to mental. Conservation of this structure involves equilibration of all its levels. Therefore, the conservation of new levels of mental organization requires their integration with other levels of the cognitive structure; one of these levels is the level of physical interaction with the environment. Conservation of the entire human cognitive structure requires the equilibration of the level of mental organization with the level that sustains physical interactions with the environment; in other words, it requires the establishment of one-to-one correspondence between mental operations and the level that sustains our physical interactions with the environment. One should point out that the establishment of one-to-one correspondence serves primarily the function of conservation, not to prove the validity of mental constructs. Even if our mental constructs may ultimately prove to be incomplete or even wrong, the fact that our mental operations are more powerful than operations that sustain our interactions with the environment, we can still establish their one-to-one correspondence, which explains why even theories that have turned out to be limited (like the geocentric system) find support in empirical facts.

The perspective that views knowledge and knowledge production through the prism of the process of creation does much more than explain what makes the different and seemingly dissonant elements of the current definition of knowledge compatible with each other. It shows how all of them are necessary and how they complement each other. This perspective also explains how knowledge growth occurs. It shows that knowledge growth is always exponential and can only be exponential due to the combinatorial nature of knowledge creation. In other words, there is no new knowledge without an exponential growth.

**Resolving the Controversy over Research Productivity**

*Indicators and Metric of Knowledge Growth*

The perspective on knowledge production presented in this article shows that the source of knowledge growth is the creation of new and increasingly more powerful levels of mental organization. Each new level and more powerful level of organization is a result of the full equilibration of all operations sustained by the level of organization from which the new level has emerged. Full equilibration means that all operations sustained by a given level of organization are in equilibrium with each other and with their opposites. Since equilibration involves multiplication of operations, not their addition, the new level is not merely a sum total of all equilibrated operations. It emerges as a result of combinations of operations, or operations on operations. Operation on operations is a more powerful form of organization since this form includes operations involved in combinations and operations that regulate these combinations. Due to multiplication of operations, the new level of organization offers more possibilities, or degrees of freedom, and, for this reason, is more powerful than the level from which it has emerged and which it regulates due to its superior combinatorial power.

The creation of new and more powerful levels of organization is what makes knowledge growth possible. Since equilibration involves multiplication of operations, the number of possible operations grows exponentially: for *n*-number of operations, the number of new possible combinations will be equal to $n^2$. Thus, knowledge can grow only exponentially. This conclusion may appear to be counterintuitive only if one does not realize that the basic unit in the evolution of knowledge is level of organization, not a specific construct, entity, or idea. The power of the new level of organization will always be exponentially greater that the power of the level from which it has emerged.

The perspective presented in this article also shows that we cannot determine research productivity based exclusively on quantitative data. Knowledge growth is first and foremost about power; it is about new possibilities, or new degrees of freedom. Numbers alone do not reflect this property. In his article Guzey shows the daunting complexity of measuring novelty by using such units of measurement as new ideas.

Ideas are mental constructs. They are products of mental operations. Mental operations are not isolated units. They are connected with other mental operations and together they constitute an ensemble that represents a level of mental organization. A level of mental organization has identifiable parameters and its unique organizing principle. The evolution of levels of organization is essentially the evolution of their organizing principle. Not individual ideas, but levels of organization mark evolutionary advances. Therefore, using levels of organization is essential for determining the dynamics of knowledge production and intellectual growth. Levels of organization are always historical and tied to evolution. Ideas can be trans historical and may not necessarily be subject to evolutionary change.

As has been pointed out, the main difference between levels of organization is exponential power differential. Each emerging level of organization sustains more operations than the level from which it has emerged. There is an easy way to determine which of the adjacent level is superior: the superior level will be always more inclusive. Thus, greater inclusiveness is one good indicator of knowledge growth. The more powerful level of organization will include all operations sustained by the level of organization from which it has emerged and operations that regulate them. The less powerful level of organization will not be able to include operations sustained by the more powerful one.

Human cognitive system includes many levels of organization. When a cognitive system changes, all levels of its organization change too. Human cognitive system is also a hierarchical structure that is regulated by the global level of organization. The global level of organization is the most general and the most powerful level in human cognitive systems. The emergence of a new global level indicates the change of the cognitive system as a whole. This global level represents the way we view reality as a whole, or what we also often call a worldview.

There are several types of global levels of organization of human cognitive systems that we know; they sustain their specific types of worldview. Deification of nature characterizes archaic civilizations and cultures. This worldview in which nature is god goes back to the beginning of human civilization. The concept of nature as agent is the organizing principle of this worldview. The second type that followed the archaic one recognized the centrality of transcendental agency, or God that transcends nature. This type is characteristic for main world religions, such as Christianity, Islam, Buddhism, and others. The third type is the one that still dominates human civilization. We call it modernity. The recognition of the pre-eminence of human agency is the characteristic feature and the main organizing principle of this worldview. Each time a new worldview emerged it marked a profound change in human practices: from organizing society to organizing production, systems of beliefs and norms, moral and aesthetic values, and much more.

These three types of worldview represent an evolutionary progression. As all other evolutions, this evolution is not and cannot be finite. The source of this evolution, as that of all other evolutions, is conservation; and conservation does not have a finite goal; or rather, the goal of conservation is within itself. Conservation does not have a beginning and does not have an end.

The emergence of a new global level of organization is inevitable. It will necessarily give rise to a new worldview—a new approach to reality defined by a new organizing principle and generating new practices. The emergence of such new worldview is a good indicator of knowledge growth.

As has been mentioned earlier, knowledge growth and economic production are integral to each other. The recognition of this fact is universal. All current economic theories agree that new knowledge and ideas generated by the human mind are the primary source of economic growth.[47] The fact that economic growth and knowledge production are connected suggests that knowledge growth must lead to increase in economic production. Since knowledge grows exponentially, so must also economic production. Consequently, an exponential economic growth is another good indicator of increased knowledge production.

This article emphasizes the role of the process of creation in the evolution of knowledge. As has been explained elsewhere,[48] the process of creation works on inclusion of differences. Combining differences is what creates new and increasingly more powerful levels of organization. Differences, therefore, are the main resource for knowledge production. If knowledge does not grow, it means that this resource is declining.

The explanation of how the process of creation works in general shows that it never runs out of differences. As differences are combined, they are conserved, not eliminated. Also, equilibration of differences inevitably produces disequilibrium; in other words, the process of creation generates new differences as it evolves. We should never run out of differences.

The decline in research productivity indicates that differences are either not utilized or are utilized efficiently. Our current social practice, including knowledge production, tends to shun and even suppress differences. The fact is well known and has produced much criticism.[49] The suppression of differences is abundant in our social practice, including science and knowledge production in general. We simply cannot have any knowledge growth as long as exclusionary practices remain dominant.

New and increasingly more powerful levels of mental organization create new possibilities and offer access to new resources. New resources offset the inevitable advance of entropy and maintain entropy production at zero. Advances in knowledge growth prevent depletion of resources. Since nature is a very important resource for humanity, knowledge growth prevents depletion of resources and the degradation of the environment. Conversely, if we fail to maintain exponential growth, we will not be able to gain access to new resources and, consequently, will have to increase our use of available resources—both material and human. The exponential increase in the use of resources will result in resource depreciation and environmental degradation.

The depletion of resources and the degradation of the environment are two major unsolved problems that our civilization faces today. Indeed, there is no lack of proposals for addressing these problems. One of the most popular solutions, particularly among environmentalists, focuses on reducing consumption and production. This approach represents merely a pragmatic knee-jerk reaction to the current conditions, not a product of a comprehensive assessment of the origin of the problem. The idea is that if we reduce consumption and production, we will stop the depletion of resources and the degradation of the environment. There is no theoretical depth to such solutions. They focus on effects rather than causes. Their proponents believe that by addressing the effects, they will treat the causes of this predicament. Nothing can be further from the truth: you cannot cure a disease by treating its symptoms. Only by addressing the causes, we can change the effects, not the other way around. However, the effects—the depletion of resources and the degradation of the environment--are also excellent indicators of the decline in knowledge production. They also correlate exponentially with the decline of knowledge growth.

The same pattern is present in the relationship between economy and population growth. People are "the ultimate resource" that we have.[50] A failure to achieve an exponential growth will result in an exponential decline in productivity, or inefficiency in using the human resource. Studies cited earlier show that if we fail to achieve an exponential growth in productivity, we will need to use more people.[51] That is what creates the momentum for population growth; and this growth follows the exponential power law, as Malthus has shown. Considering Julian Simon's point that humanity itself is "the ultimate resource," the conclusion that a stagnant population is the ultimate resource crisis is quite appropriate.[52]

This is to argue that the cause of population growth is inefficient production. Current proposals to limit this growth will not succeed because they do not address the main cause of the population growth—exponential increase in production inefficiency. All efforts to limit the global level of population have failed. There have been numerous summits and conferences over the last several decades that discussed this issue.[53] The only result that they produced, as critics charge, were numerous statements to the effect that "reducing population growth was a necessary part of ecologically sustainable development."[54] This singular lack of success has caused much criticism of the entire population science. George Weigel, president of the Ethics and Public Policy Center in Washington, D.C.—one of the harshest critics—has charged that population science has little merit and only produces myths, miscalculations, and bogus predictions.[55]

The fact is that the population growth is not an independent variable.[56] Its source is production inefficiency that grows exponentially and requires exponential compensations. The continued acceptance of child labor in the developing world as a partial solution to poverty speaks to the shortage of labor due to low productivity.[57] Efforts to use compulsion to reduce population are and will continue to be extremely unpopular and utterly ineffective. They will simply put more pressure on the economy that will be unable to maintain its current level of economic production. The conclusion is certainly counterintuitive, but the logic of population growth suggests that an efficient use of human resources requires exponential growth in productivity. Such growth is the only way to avoid overpopulation.

To conclude, the perspective that that uses the process of creation as its main organizing principle offers a new approach toward assessing knowledge production. Rather than using micro data on specific innovation and new ideas in different spheres of research and in individual industries, the new approach focuses on global indicators of knowledge growth: the emergence of new global levels of regulation and the effects of knowledge production on economic growth, population dynamics, and the use of resources. This approach also offers a simplified exponential metric in assessing knowledge production.

*Assessing the Current State of Research Productivity*

The discussion of the controversy over research productivity shows that the approaches proposed by the participants do not resolve this controversy. As Guzey summarized in his contribution, the central question in this debate—whether research productivity is in decline or not—remains unanswered. Guzey points out that the very data used for assessing research productivity does not allow making a definitive conclusion. We still do not know what the actual state of research productivity is one way or another.

The approach for assessing research productivity outlined in this article focuses on a different set of indicators. Rather than using micro data from specific industries and research areas, its data relates to macro dimensions: the general features of the process that generates knowledge and the effects of knowledge growth on economic production, use of resources, and population changes.

According to the approach presented in this article, one important global indicator of knowledge growth is changes in the global level of organization that regulates the human cognitive system. This is the level that sustains our general view of reality, or what we often call worldview.

The view of reality that currently dominates practically all research areas is not much different from what it has been over the past few centuries. Despite variations and modifications, it still remains embedded in anthropocentrism. This view of reality critically depends on mental constructs created by humans that are inevitably subjective and arbitrary, despite claims of objectivity. Humanity and human biases define this worldview. The persistence of the human-centeredness is a good indicator that knowledge is not growing and, consequently, is in decline. As this article has argued, knowledge production cannot be static: it must either grow or decline.

There is a general agreement among researchers that knowledge production is critical for economic progress. As this article has shown, the rate of knowledge growth can only be exponential. Consequently, knowledge growth should generate exponential increase in economic production. The fact that today's economy does not demonstrate exponential growth is another indicator that knowledge production is in decline.

This article has also stressed that knowledge growth is our safeguard against depletion of resources, both human and non-human. Knowledge growth offsets unavoidable dissipative entropic processes. It makes possible to maintain a zero-level of entropy production. The current picture is one of exponential depletion of resources; its most vivid evidence is the degradation of the environment: nature is no longer capable of recycling human waste that is accumulating in Earth's soil, water, and atmosphere at the rate that poses a serious threat to human survival on the planet. This rapid increase in the amount of human waste in the terrestrial environment is also an indicator of the decline in knowledge production.

Finally, as this article has pointed out, knowledge advances save labor. A decline of knowledge production inevitably increases the demand for labor, thus creating pressure for population growth. Back in the 18$^{th}$ century Thomas Malthus pointed out the tendency of population to grow exponentially, far in excess of the capacity of economic production to sustain increasing number of people. Unlike Malthus, this article does not see this tendency as inevitable. It sees the current growth of population as the necessary consequence of the decline in knowledge growth and the increasing demand for labor that this decline causes. As this article concludes, the current and seemingly unstoppable tendency for population growth is a clear indication of the decline in knowledge production.

These general indicators are clear and are supported by factual evidence. They are due to the very nature of the process of creation that fuels knowledge growth. Unlike the indicators used in the current debates, the use of these indicators does not rely on mental constructs created by humans. They are not a subject of controversy or disagreement. The exponential index of these indicators additionally points to their common source.

All the current indicators demonstrate what many researchers conclude on the basis of intuition and selective data: the current knowledge production is definitely in a downward spiral. This conclusion resolves the current controversy over research productivity.

**Enhancing Research Productivity**

The decline in knowledge production is not a new phenomenon. There have been several periods in human history when knowledge stagnated. For example, numerous researchers have registered a marked decline of science in Europe after the demise of Antiquity. Although some intellectual stirrings occurred during the late Middle Ages and the Renaissance, no significant advances in knowledge that rose to the level of achievement that characterized Antiquity took place before the onset of modernity. Only during the Industrial Age Europe began to experience a major and consistent upswing in science, technology, and other intellectual spheres.

The pattern of knowledge production has so far been cyclical. Periods of expansion and contraction of knowledge followed each other. Even the Modern Age that is characterized by fairly stable knowledge growth reveals periodic cyclical patterns in the evolution of knowledge. There was, for example, a rapid increase in scientific knowledge in the $19^{th}$ century that slowed down by the mid-$20^{th}$ century. As this article concludes, this contraction continues to this day, despite occasional spikes in some areas. For example, no major breakthroughs have occurred in theoretical physics since the formulation of quantum mechanics and theory of relativity.

Thomas Kuhn provides a detailed discussion of this cyclical tendency in his now famous book *The Structure of Scientific Revolutions*.[58] Kuhn's book is more descriptive than explanatory. He points to the existence of cycles but does not really explain their origin. Kuhn attributes the cyclicity in the evolution of knowledge to some natural tendency and leaves it at that. Knowledge advances result from natural attrition of the practitioners of what Kuhn calls "normative science," or, as Kuhn succinctly remarks, when "they die out."

The perspective presented in this article makes possible to explain the cyclicity in knowledge production. The explanation pivots on the simple relationship between growth and resources. Declines in knowledge growth are due to an inefficient use of resources. The current decline in knowledge production is, in this sense, no different from others that occurred in the past.

As has been explained earlier, the process of creation involves equilibration, which is a form of integration, and the production of disequilibrium, or differentiation. Equilibration integrates individual operations, or subsystems. Such integration gives rise to a new and more powerful global level of organization. The equilibration of the new global level with the level that sustains local interactions from which it has emerged leads to differentiation of the global level. This differentiation requires re-equilibration. Both integration and differentiation play the principal role in the process of creation. Differences are the essential resource for the process of creation and, consequently, for the production of knowledge. The process of creation that operates in nature never runs out of this resource because as it "consumes" differences, it also generates differences. There is never a shortage of differences in nature as integration and differentiation are always in balance with each other. As a result, the process of creation does not lapse.

Why then human cognitive systems have lapses in their evolution? Is there a shortage of differences? If so, what is the cause of such shortage? The answer is simple. We tend to shun, exclude, and suppress differences. The social practice that is dominant in our civilization is exclusionary.[59] This suppression of differences is equivalent in its effect to "depletion" of this important resource.

Some may object by pointing out that human social practice involves both inclusion and exclusion. The examples of inclusion, even if non-negligible, do not change the fact that the practice is exclusionary. The inclusion we practice is selective. As has been argued elsewhere, selective inclusion is basically a form of exclusion.[60] The fact is that the function of inclusion is the creation of new and increasingly more powerful levels of organization. If some operations/subsystems are excluded, they are not conserved; and without their conservation, the rise of a new and more powerful level of organization is impossible. Even if one operation, or sub-system, is excluded, the cumulative effect on the entire will be non-conservation; in other words, the entire system will not be conserved and advance will not take place. Exclusion and inclusion are totalizing practices. They are absolute and absolutely incompatible with each other. As opposites, they are not equivalent to each other. Inclusion is constructive. It creates new levels of organization and makes evolution possible. Finally, it is associated with the fundamental processes of conservation and creation that sustain our universe. By contrast, exclusion is a result of the absence of inclusion. It has no autonomous source and is ultimately destructive.

But what are the reasons for exclusion? Why do we exclude differences?

The most obvious function of exclusion is protection. By excluding some differences, we try to protect differences that we represent; in knowledge production we try to protect our own mental constructs. However, such reaction to differences reflects a total lack of understanding of how conservation works. Conservation works on inclusion and the creation of new and more powerful level of organization, not on exclusion. We do not conserve our constructs by shunning and suppressing differences. Obviously, the fact that we widely use exclusionary practices shows that we do not embrace and, consequently, do not understand the process of creation and the way it works. That is probably the most important reason why we exclude differences. We simply do not understand the important role of differences in conserving our own mental constructs. As a result, we view differences as a threat to our own vision of reality and the order this vision represents for us.

Indeed, we recognize today that our view of reality is our creation and that it is only an approximation to reality. The admission creates a problem. However, rather than pursue the solution of this problem, we simply proclaim it irresolvable. The suppression of differences continues. Naked political power is the sole referee in matters related to knowledge and truth. Those who use this practice do not even pretend that they stand for truth; moreover, they claim that truth is in principle unattainable.

Since decline in knowledge production is a result of suppression of differences, we can prevent such declines by embracing the practice of inclusion of differences. Contrary to Kuhn's argument, there is nothing inevitable about downward spirals in knowledge growth. It is simply a result of the profoundly flawed practice. We can eliminate this flaw by abandoning our current practice, by accepting the centrality of the process of creation, and establishing a new practice that uses the process of creation as its central organizing principle. The introduction of the new practice would necessarily involve the abandonment of our current worldview rooted in anthropocentrism.

Shifting from one worldview to another has never been easy; but it is possible and, as history shows, necessary. There have been several such major shifts in the past. Deification of nature was the main organizing principle in the way that ancient civilizations viewed reality. The extant indigenous cultures in the world today still hold on to this view.

All major world religions and the cultures they shaped had abandoned worldviews that centered on deification of nature. Instead, they adopted a new worldview that used transcendental agency as its main organizing principle. The new worldview did not completely abandon the archaic approach but adapted and integrated in into the new vision. Christianity with its multiple denominations and sects was the European version of this transition.

The recognition of human agency became the organizing principle in the new worldview that marked the beginning of what we call the Modern Age, or modernity. The late Middle Ages and the Renaissance witnessed the emergence of this new worldview in Europe but the rise of similar worldviews has become a worldwide phenomenon. The story of modernity is a subject that has been thoroughly discussed; the main landmarks of modernity—industrialization, the emergence of modern science, the rise of new moral and aesthetic theories, technological advances, rising living standard and quality of life, and much more—are well known.

Since its appearance on the world stage centuries ago modernity has dominated our civilization and in some sense still continues to be dominant. However, there are numerous indications that its grip on human civilization has loosened and its influence is diminishing. The demise of modernity is not a new development. Among its early signs was the rise of the view of reality as ultimately uncertain and chaotic. Post-modernism has sealed the fate of modernism by proclaiming that reality is ultimately unknown.

There are many signs today that we need a major shift in the way we view reality and in our social practices. The decline in research productivity—a problem that remains unsolved—is but one of them. Many economists, theorists and practitioners of management, business leaders, and politicians see the need for an entirely new approach to boost economic production and solve numerous problems created by our lethargic economy. The environmental movement sees the culture of modernity as the main source of the degradation of nature and a threat to the survival of humanity. Calls for ending the exclusionary practice that still prevails in many spheres of our life have become very common these days.

There is nothing sacred about the view of reality that is associated with modernity. Shifts of such magnitude have occurred in the past and they have greatly benefited humanity. Where would we be today without these shifts? Still venerating the divine agency of nature? In fact, one can observe such reversions in the way that modern day environmentalists extol what they see as the "superior wisdom" of archaic cultures. This going back to the past is not a solution; it is an illusion. Our civilization cannot turn back. The shift must take place and it will.

There is no shortage of competing visions for the world's future. The rivalry in this competition is fierce and is quickly reaching crescendo. None of the alternative visions is in the position to prevail. Their standoff is a source of tensions and divisions, which became a problem in their own right. The solution of this problem is in finding the approach that would reconcile all these different visions. The compromise among them is not a possibility. This solution must rest on something that they all value and share.

The only feature that all these competing visions have in common is the fact that all of them are created. Therefore, there is a process that has created them. This process can serve as the foundation for an all-inclusive worldview that will encompasses all these competing visions

As this article argues, the process of creation transcends the human mind and its constructs. This process is not a result of human creation; rather, it has given rise to humanity. Because of its transcendent and all-inclusive nature, this process is the only candidate to become the organizing principle for a new worldview that would include all current proposals as its particular cases—that is, cases that are true under specific conditions or assumptions.

In one important respect, though, the shift to the new worldview will be different from its predecessors. All previous worldviews were essentially anthropocentric. All of them viewed reality through the prism of mental constructs created by humans. The new and growing realization is that anthropocentrism represents a subjective, incomplete, and ultimately arbitrary perspective on reality. The position of the environmentalists is highly indicative in this regard: they see anthropocentrism as the main cause of the abuse of nature by humans and call for abandoning it. The new worldview will not merely replace modernity and its central focus on humanity. It goes much deeper. The new worldview replaces the very organizing principle that defined all previous worldviews; it does not destroy those worldviews but integrates them as its particular cases.

By contrast with the anthropocentric approach, the approach centered on the process of creation is objective and universally inclusive. It includes all possible levels of organization—those that were created in the past, those that are being created now, and those that will be created in the future. In other words, like the process of creation, the new worldview will not be selective in preferring some mental constructs to others. The social practice sustained by this worldview will be universally inclusive and empowering. This new practice will ensure an uninterrupted and exponential knowledge growth.

Humans inherited the process of creation in the course of the evolution. Since the process of creation is not a human construct, the worldview that is organized around this process will not be anthropocentric. Our knowledge production practice based on this new worldview will make objective and unbiased observations of reality possible.

Moreover, the new approach makes possible to observe the process itself without falling into an infinite regress.[61]

An understanding of how this process of creation works will help us realize the true importance and significance of differences and inclusion. We will no longer fear differences as a threat, but rather view them as opportunities to create new and increasingly more powerful levels of organization that will give rise to new ideas, theories, and approaches. We will also be able to control and use the infinite power of the process of creation and use it as the infinite source of differences that will provide an inexhaustible supply of resources, both human and non-human, and thus ensure an uninterrupted, infinite, and exponential growth of knowledge.

**Conclusion**

The importance of knowledge for the survival of our civilization has never been more important than it is today; and this importance will only grow in the future. Knowledge production is the subject that is relevant to many areas of our life: from science and technology to welfare of our citizens, stability of our society, economic prosperity, the health of our planet, and much more. Problems related to knowledge production are these days a major preoccupation in public discussions that involve broad segments of the population.

Yet, as important of this subject is, the simple fact remains that we still do not know much about it. As current contributions on the subject show, we do not even know for sure what the current state of knowledge production is; we still do not have a reliable set of indicators and metric that can provide a definitive resolution to this most basic issue.

This fundamental gap in our knowledge about knowledge is not for a lack of trying or a shortage of facts. As this article argues, the main reason for our failure is that we continue to approach this subject from perspectives that are limited, subjective, and ultimately arbitrary. This article has argued for an objective approach that will not view reality through the prism of our subjective constructs. The strength of the perspective outlined in this article is that it abandons the anthropocentric focus that refracts reality through the prism of mental constructs that we have created. The distinctive feature of the new prospective is that it uses the process of creation that fuels knowledge production as its main organizing principle. It focuses on major global characteristics of this process and the general effects of knowledge production on related areas, such as economic progress, use of resources, and population growth.

This approach provides a definitive answer to the question regarding the state of our knowledge production. According to the indicators and metric used in this approach, there can be no doubt that the rate of our knowledge growth is definitely in decline, despite some successes in specific areas. The most important telltale sign is the fact that the global level of mental organization that regulates our entire cognitive system remains unchanged in its basic features since at least the last period that witnessed significant knowledge growth over two centuries ago.

Other important signs include the lack of exponential economic growth, the continued depletion of resources and the degradation of the environment, and the exponential population growth. The exponential metric is of all these signs indicates their intrinsic connection to knowledge growth. As this article has argued, knowledge growth can only be exponential and its effects also bear an exponential mark.

The establishment of the connection between knowledge growth and the process of creation is an important contribution of this article. An understanding of the process of creation makes possible to differentiate specific stages in knowledge production and offers specific how-to recommendations to raise research productivity. The understanding of the process of creation provided in this article reveals the vital link between production of knowledge and the fundamental processes that sustain our universe: conservation, creation, and evolution. By embracing the process of creation as the main organizing principle of our knowledge practice, we will be able to understand how our knowledge evolves. The new practice will eliminate the gap that has precluded and still precludes us from understanding this process—the gap that hindered the evolution of our knowledge since the beginning of human civilization.

The practice that is based on the new perspective will solve the problem with research productivity we face today. Moreover, relying on this perspective will ensure that the problem with research productivity does not occur in the future. The institutionalization of the new practice will work against any disruptions in the process of creation and will ensure exponential growth of knowledge. Inclusion will be central to this new practice; it will be absolute and universal--no exceptions. The practice based on universal inclusion will result in an infinite, uninterrupted, and exponential knowledge growth. It will not only solve the problems of knowledge growth for now, but also for all times to come.

# ENDNOTES

[11] Nicholas Bloom, Charles I. Jones, John Van Reenen, and Michael Webb, "Are Ideas Getting Harder to Find?" Working Paper. Working Paper Series. National Bureau of Economic Research, September 2017, p. 46, https://doi.org/10.3386/w23782.

[12] Moore's law is a term associated with the observation made by Gordon Moore in 1965 that the number of transistors in a dense integrated circuit (IC) doubles about every two years.

[13] Bloom, et al., "Are Ideas Getting Harder to Find?" p. 46.

[14] "Exponential Growth Is the Baseline," *The Roots of Progress*, February 21, 2021. https://rootsofprogress.org/exponential-growth-is-the-baseline; Böing and Hünermund, "A Global Decline in Research Productivity? p. 1; Derek Thompson, "America Is Running on Fumes," *The Atlantic*, December 1, 2021 https://www.theatlantic.com/ideas/archive/2021/12/america-innovation-film-science-business/620858/; Tom Relihan, "New Ideas Are Getting Harder to Find—and More Expensive," *MIT Sloan*, August 31, 2018, https://mitsloan.mit.edu/ideas-made-to-matter/new-ideas-are-getting-harder-to-find-and-more-expensive; Tsutomu Miyagawa, and Takayuki Ishikawa, "On the Decline of R&D Efficiency," *VoxEU.Org* (blog), November 13, 2019, https://voxeu.org/article/decline-rd-efficiency.

[15] Alexey Guzey, "Issues with Bloom et al's 'Are Ideas Getting Harder to Find?' And Why Total Factor Productivity Should Never Be Used as a Measure of Innovation," https://guzey.com/economics/bloom/ (accessed January 31, 2022).

[16] Böing and Hünermund, "A Global Decline in Research Productivity."

[17] Benjamin F. Jones, "The Burden of Knowledge and the 'Death of the Renaissance Man': Is Innovation Getting Harder?" Working Paper. Working Paper Series. National Bureau of Economic Research, May 2005, https://doi.org/10.3386/w11360.

[18] Jones, "The Burden of Knowledge and the 'Death of the Renaissance Man': Is Innovation Getting Harder?" Working Paper. Working Paper Series. National Bureau of Economic Research, May 2005, p. 28, https://doi.org/10.3386/w11360.

[19] Christian William Callaghan, "Growth Contributions of Technological Change: Is There a Burden of Knowledge Effect?" *Technological Forecasting and Social Change* 172 (November 1, 2021): 121076, p. 3, https://doi.org/10.1016/j.techfore.2021.121076..

[20] Salman Yousuf Guraya, Khalid Ibrahim Khoshhal, Muhamad Saiful Bahri Yusoff, and Maroof Aziz Khan, "Why Research Productivity of Medical Faculty Declines after Attaining Professor Rank? A Multi-Center Study from Saudi Arabia, Malaysia and Pakistan," *Medical Teacher*, vol. 40, no. 1 (July 6, 2018), pp. S83–89, https://doi.org/10.1080/0142159X.2018.1465532.

[21] Flora F. Tien and Robert T. Blackburn, "Faculty Rank System, Research Motivation, and Faculty Research Productivity: Measure Refinement and Theory Testing," *The Journal of Higher Education*, vol. 67, no. 1 (January 1, 1996), pp. 2–22, https://doi.org/10.1080/00221546.1996.11780246.

[22] Corinne A. Kratz, "Ideologies of Access and the Politics of Knowledge Production," *Men & Masculinities*, vol. 11, no. 2 (December 2008), pp. 193–200; Morton White, "The Politics of Epistemology," *Ethics*, vol. 100, no. 1 (October 1, 1989), pp. 77–92.

[23] Nathalie Lazaric and Alain Raybaut, "Knowledge Creation Facing Hierarchy: The Dynamics of Groups inside the Firm," *Journal of Artificial Societies and Social Simulation*, vol. 7, no. 2 (2004), http://jasss.soc.surrey.ac.uk/7/2/3.html.

[24] Sharon G. Levin and Paula E. Stephan, "Research Productivity Over the Life Cycle: Evidence for Academic Scientists," *The American Economic Review,* vol. 81, no. 1 (1991), pp. 114–32.

[25] Timothy Salthouse, "Consequences of Age-Related Cognitive Declines," *Annual Review of Psychology*, vol. 63, no. 1 (2012), pp. 201–26, https://doi.org/10.1146/annurev-psych-120710-100328.

[26] Dean Keith Simonton, "Age and Outstanding Achievement: What Do We Know after a Century of Research? – PsycNET," https://psycnet.apa.org/doiLanding?doi=10.1037%2F0033-2909.104.2.251 (accessed January 27, 2022); Wolfgang Stroebe, "The Graying of Academia: Will It Reduce Scientific Productivity?" *The American Psychologist*, vol. 65, no. 7 (October 2010), pp. 660–73, https://doi.org/10.1037/a0021086.

[27] Lisa Marshall, "Does Faculty Productivity Really Decline with Age? New Study Says No." *CU Boulder Today*, October 17, 2017, https://www.colorado.edu/today/2017/10/17/does-faculty-productivity-really-decline-age-new-study-says-no.

[28] Laura A. Hunter and Erin Leahey, "Parenting and Research Productivity: New Evidence and Methods," *Social Studies of Science*, vol. 40, no. 3 (June 1, 2010), pp. 433–51, https://doi.org/10.1177/0306312709358472.

[29] Kabir Sehgal and Deepak Chopra, "Stanford Professor: Working This Many Hours a Week Is Basically Pointless. Here's How to Get More Done—by Doing Less," *CNBC*, March 20, 2019, https://www.cnbc.com/2019/03/20/stanford-study-longer-hours-doesnt-make-you-more-productive-heres-how-to-get-more-done-by-doing-less.html.

[30] Jyoti Madhusoodanan, "The Pandemic's Slowing of Research Productivity May Last Years—Especially for Women and Parents," *Science*, October 26, 2021, https://www.science.org/content/article/pandemic-s-slowing-research-productivity-may-

[38] Ettore Bolisani and Constantin Bratianu, "The Elusive Definition of Knowledge," *Knowledge Management and Organizational Learning*, 2018, pp. 1–22, https://doi.org/10.1007/978-3-319-60657-6_1.

[39] Paul K. Moser, *The Theory of Knowledge:  A Thematic Introduction* (New York:  Oxford University Press, 1998); Louis P. Pojman, *The Theory of Knowledge:  Classical and Contemporary Readings* (Belmont, CA:  Wadsworth Pub, 1999); Bolisani and Bratianu, *The Elusive Definition of Knowledge*; Claudio F. Costa, "A Perspectival Definition of Knowledge," *Ratio*, vol. 23, no. 2 (June 2010), pp. 151–67, https://doi.org/10.1111/j.1467-9329.2010.00458.x;  Tommi Vehkavaara, "Extended Concept of Knowledge," http://www.uta.fi/~attove/vehka-f.htm (accessed November 8, 2011); E. J. Coffman, "Is Justified Belief Knowledge?  Critical Notice of Jonathan Sutton, without Justification," *Philosophical Books*, vol. 51, no. 1 (January 2010), pp. 1–21, https://doi.org/10.1111/j.1468-0149.2010.00498.x; David N. Mermin, "Whose Knowledge?" *ArXiv:Quant-Ph/0107151*, July 30, 2001. http://arxiv.org/abs/quant-ph/0107151.

[40] Moser, *The Theory of Knowledge,* p. 14.

[41] Moser, *The Theory of Knowledge*, p. 15.

[42] Louis P. Pojman, *The Theory of Knowledge:  Classical and Contemporary Readings*. 2nd ed. Belmont, CA: Wadsworth Pub, 1999, p. 2; Coffman, "Is Justified Belief Knowledge?"

[43] Eshel Ben Jacob, Yoash Shapira, and Alfred I. Tauber, "Seeking the Foundations of Cognition in Bacteria:  From Schrödinger's Negative Entropy to Latent Information," *Physica A:  Statistical Mechanics and Its Applications*, vol. 359 (January 1, 2006), pp. 495–524, https://doi.org/10.1016/j.physa.2005.05.096.

[44] Gennady Shkliarevsky, "Conservation, Creation, and Evolution: Revising the Darwinian Project," *Journal of Evolutionary Science*, vol. 1, no. 2 (September 25, 2019), pp. 1–30, https://doi.org/10.14302/issn.2689-4602.jes-19-2990.

[45] Jean Piaget, *The Origins of Intelligence in Children* (Madison, Conn.:  International Universities Press, Inc., 1998).

[46] Gennady Shkliarevsky, *The Civilization at a Crossroads:  Constructing the Paradigm Shift* (Raleigh, NC:  Glasstree Publishing, 2017), https://www.researchgate.net/publication/318431832_The_Civilization_at_a_Crossroads_Constructing_the_Paradigm_Shift.

[47] Andrea Bassanini, Stefano Scarpetta, and Ignazio Visco, "Knowledge, Technology and Economic Growth:  Recent Evidence from OECD Countries," *SSRN Scholarly Paper*. (Rochester, NY: Social Science Research Network, October 1, 2000),

https://doi.org/10.2139/ssrn.246375; Matti Pohjola, ed., *Information Technology and Economic Growth: A Cross-Country Analysis*. Working Papers WP, 2000. https://doi.org/10.22004/ag.econ.295500; Philoppe Aghion and Peter Howitt, "Some Thoughts on Capital Accumulation, Innovation, and Growth," *Annals of Economics and Statistics*, no. 125/126 (2017), pp. 57–78. https://doi.org/10.15609/annaeconstat2009.125-126.0057; Jason Crawford, "Exponential Growth Is the Baseline," *The Roots of Progress*, February 21, 2021, https://rootsofprogress.org/exponential-growth-is-the-baseline.

[48] Gennady Shkliarevsky, "In Quest for Justice: Solving the Problem of Inclusion and Equality) (June 8, 2021), *SSRN*, https://ssrn.com/abstract=3862630 or http://dx.doi.org/10.2139/ssrn.3862630

[49] Gennady Shkliarevsky, "In Quest for Justice: Solving the Problem of Inclusion and Equality) (June 8, 2021), *SSRN*, https://ssrn.com/abstract=3862630 or http://dx.doi.org/10.2139/ssrn.3862630

[50] "Exponential Growth Is the Baseline," *The Roots of Progress*, February 21, 2021. https://rootsofprogress.org/exponential-growth-is-the-baseline.

[51] Bloom, et al., "Are Ideas Getting Harder to Find?"

[52] "Exponential Growth Is the Baseline," February 21, 2021, *The Roots of Progress*, https://rootsofprogress.org/exponential-growth-is-the-baseline.

[53] Population Summit (1993 : New Delhi, India), and Francis Graham-Smith, *Population, the Complex Reality*; "Science Summit on World Population: A Joint Statement."

[54] S. Ryan Johansson, "Complexity, Morality, and Policy at the Population Summit," *Population and Development Review*, vol. 21, no. 2 (1995), pp. 361–86, p. 361, https://doi.org/10.2307/2137499.

[55] George Weigle, "False Precision and Population Science," *Baltimore Sun*, September 7, 1994, https://www.baltimoresun.com/news/bs-xpm-1994-09-07-1994250032-story.html.

[56] Melissa Mayer, "What Are Environmental Problems Due to Population Growth?" *Sciencing*, 2018. https://sciencing.com/environmental-problems-due-population-growth-8337820.html.

[57] Johansson, "Complexity, Morality, and Policy at the Population Summit," p. 382.

[58] Thomas Samuel Kuhn, *The Structure of Scientific Revolutions* (Chicago: University of Chicago Press, 2012).
28

[59] Shkliarevsky, "In Quest for Justice"; Gennady Shkliarevsky, "Confronting the Aporias of the Inclusion Practice" (May 20, 2019), *SSRN*, https://ssrn.com/abstract=3423644 or http://dx.doi.org/10.2139/ssrn.3423644

[60] Shkliarevsky, "Confronting the Aporias of the Inclusion Practice."

[61] Niklas Luhmann, *Social Systems* (Stanford: Stanford University Press, 1995), p. 479; Gennady Shkliarevsky, "The Paradox of Observing, Autopoiesis, and the Future of Social Sciences," *Systems Research and Behavioral Science*, vol. 24, no. 3 (2007), pp. 323–32, https://doi.org/10.1002/sres.811.

Simonton, Dean Keith. "Age and Outstanding Achievement: What Do We Know after a Century of Research? – PsycNET," https://psycnet.apa.org/doiLanding?doi=10.1037%2F0033-2909.104.2.251 (accessed January 27, 2022).

Stroebe, Wolfgang. "The Graying of Academia: Will It Reduce Scientific Productivity?" *The American Psychologist*, vol. 65, no. 7 (October 2010), pp. 660–73, https://doi.org/10.1037/a0021086.

Thompson, Derek. "America Is Running on Fumes." *The Atlantic*, December 1, 2021 https://www.theatlantic.com/ideas/archive/2021/12/america-innovation-film-science-business/620858/.

Tien, Flora F., and Robert T. Blackburn. "Faculty Rank System, Research Motivation, and Faculty Research Productivity: Measure Refinement and Theory Testing." *The Journal of Higher Education*, vol. 67, no. 1 (January 1, 1996), pp. 2–22, https://doi.org/10.1080/00221546.1996.11780246.

Vehkavaara, Tommi. "Extended Concept of Knowledge," http://www.uta.fi/~attove/vehka-f.htm (accessed November 8, 2011);

Vollrath, Dietrich E. "Yes, Ideas Are Harder to Find. Don't Panic Yet," January 24, 2017, https://growthecon.com/blog/Idea-TFP/.

Weigle, George. "False Precision and Population Science." *Baltimore Sun*, September 7, 1994, https://www.baltimoresun.com/news/bs-xpm-1994-09-07-1994250032-story.html.

White, Morton. "The Politics of Epistemology." *Ethics*, vol. 100, no. 1 (October 1, 1989), pp. 77–92.

Williams, Ray. "The Decline of Productivity and How To Fix It." *Ray Williams* (blog), June 8, 2018, https://raywilliams.ca/the-decline-of-productivity-and-how-to-fix-it/;

Wladawsky-Berger, Irving. "In an Era of Tech Innovation, Whispers of Declining Research Productivity." *The Wall Street Journal*, July 13, 2018, https://www.wsj.com/articles/in-an-era-of-tech-innovation-whispers-of-declining-research-productivity-1531500122

35